\documentclass[aps,prl,twocolumn,showpacs,preprintnumbers,amsmath,amssymb,superscriptaddress,notitlepage]{revtex4-2}

\usepackage[colorlinks=true,allcolors=blue]{hyperref}
\usepackage{xcolor}
\usepackage{txfonts}
\usepackage{graphicx}
\usepackage{color}
\usepackage{dcolumn}
\usepackage{bm}
\usepackage{amsmath}
\usepackage{feynmp}
\usepackage{braket}
\usepackage{ulem}
\usepackage[multiple]{footmisc}
\usepackage{cleveref}
\newcommand{\chem}[1]{\ensuremath{\mathrm{#1}}}

\newcommand{\nc}{\newcommand}
\nc{\bea}{\begin{eqnarray}}
\nc{\eea}{\end{eqnarray}}
\nc{\vx}{{\mathbf{r}}}
\nc{\vm}{\mathbf{m}}

\begin{document}

\title{Asymmetric Antibimerons: Statics and Dynamics}

\author{Pavel~A.~Vorobyev}
\affiliation{
School of Physics, The University of New South Wales, Sydney 2052, Australia
}
\author{Daichi~Kurebayashi}
\affiliation{
	School of Physics, The University of New South Wales, Sydney 2052, Australia
}
\author{Oleg~A.~Tretiakov}
\email{o.tretiakov@unsw.edu.au}
\affiliation{
School of Physics, The University of New South Wales, Sydney 2052, Australia
}
\date{October 14, 2024}

\begin{abstract}
Nontrivial topological spin textures, such as magnetic skyrmions, are of great interest due to their potential use as information carriers in spintronic memory and logic. Here, we theoretically predict an asymmetric antibimeron (AAB) -- a topological texture in an in-plane magnetized chiral ferromagnet with $D_{2d}$ symmetry, which has yet to be observed experimentally. We show that AABs can be stabilized by an anisotropic interfacial Dzyaloshinskii-Moriya interaction, characteristic of materials with $D_{2d}$ symmetry. Using energy considerations, we explain the origin of its asymmetric shape, composed of an antivortex and a crescent-shaped vortex with opposite core polarizations. Furthermore, we demonstrate that AABs of opposite topological charge can coexist within the same ferromagnetic film, in contrast to skyrmions in out-of-plane magnetized systems. Employing micromagnetic simulations and an analytical approach based on an extension of Thiele’s equation to a system of two elliptic merons comprising the AAB, we are able to describe the current-driven dynamics of AABs. In particular, we show that AAB collisions enable controlled manipulation of the system’s topological charge. Our findings shed light on fundamental understanding of asymmetric topological magnetic solitons and provide a roadmap for their experimental observation in in-plane magnetized ferromagnets. 
\end{abstract}

\maketitle

{\it Introduction.--} Over the past few decades, various noncollinear magnetic textures have been the focus of intense research~\cite{KOSEVICH1990, GobelPhysRep2021}. 
In the vast majority of recently studied magnetic materials, the stability of these textures arises from the Dzyaloshinskii-Moriya interaction (DMI)~\cite{Dzyaloshinsky1958, Moriya1960}, which is present in crystals with broken inversion symmetry and favors perpendicular alignment of magnetic moments on neighboring sites. Magnetic skyrmions -- tiny whirls in the magnetic order with nonzero topological charge -- serve as a prominent example of such textures~\cite{Bogdanov1989, Robler2006, Nagaosa2013, Muhlbauer2009}. Their topological protection, high mobility under spin-polarized currents, and nanometer-scale size make them promising candidates for ultra-dense information storage~\cite{Romming2013, Kiselev2011, Sampaio2013}. In standard binary logic, the presence or absence of a skyrmion has been proposed to encode information bits ‘‘1'’ or ‘‘0'’~\cite{Fert2013}.

\begin{figure}[t]
	\centering
	\includegraphics[width=1\linewidth]{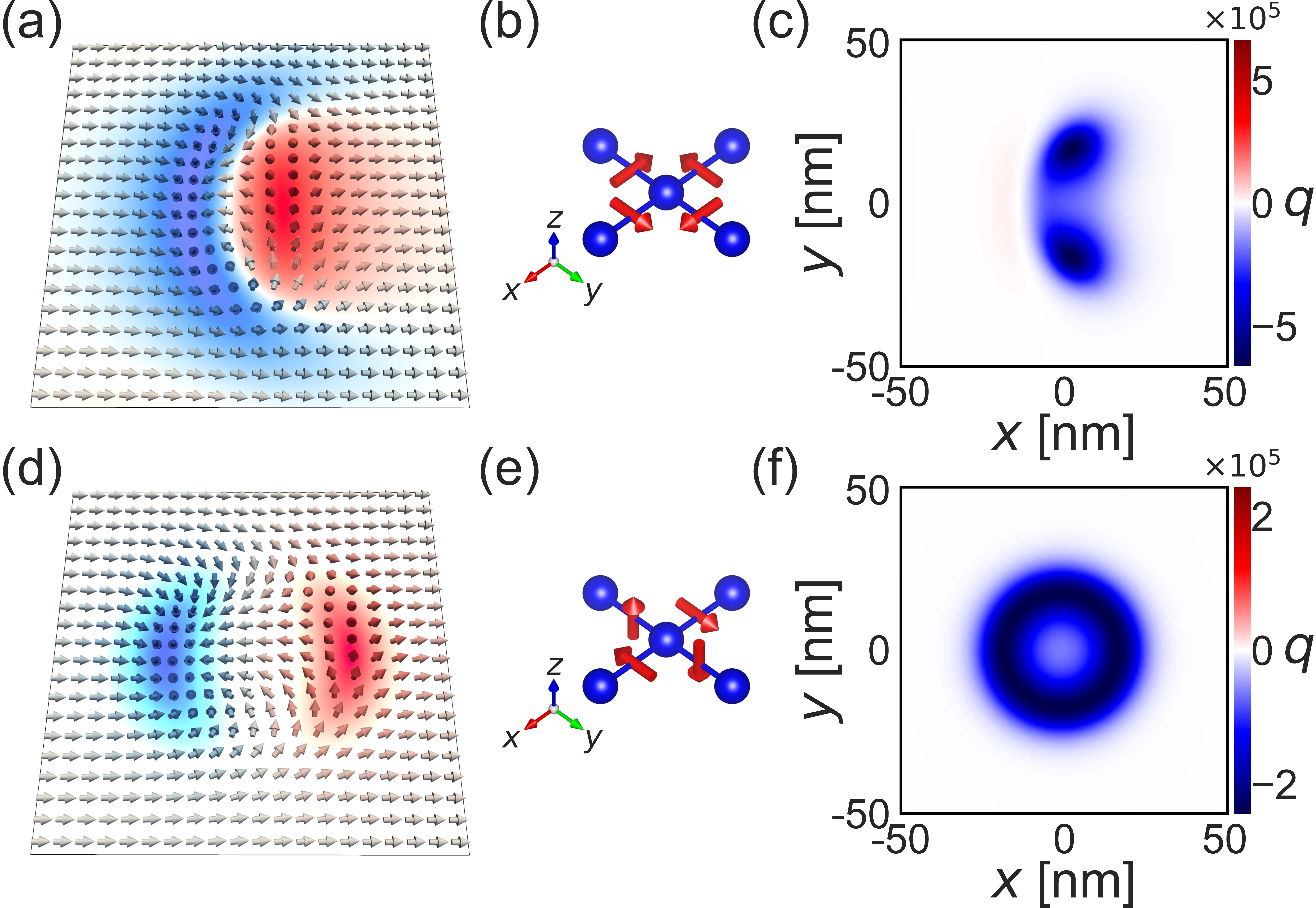}
	\caption{
		Asymmetric antibimeron: (a) Magnetic configuration, (b) DMI vectors (red arrows) between neighboring magnetic atoms, and (c) topological charge density.
		Symmetric bimeron: (d) Magnetic configuration, (e) DMI vectors, and (f) topological charge density.
	}
	\label{fig1}
\end{figure}

Nevertheless, substantial challenges remain for memory and logic applications of skyrmions. 
In particular, ferromagnetic (FM) skyrmions experience transverse deflection from the motion along the current due to the skyrmion Hall effect~\cite{Litzius2017, Jiang2017}, which can lead to their annihilation at the racetrack edges. Moreover, interference between FM skyrmion-based racetracks caused by stray fields hinders the high-density integration of three-dimensional racetrack devices \cite{Parkin2022}. One possible approach to overcoming these challenges in FM materials is to utilize in-plane counterparts of skyrmions -- symmetric bimerons~\cite{Gobel2019, Kharkov2017} shown in Fig.~\ref{fig1}(d). While skyrmions and bimerons are topologically equivalent, bimerons exhibit distinct static and dynamic properties~\cite{Li20, Shen2020, Jani21, Udalov2021, Amin2023, Leonov2017, ShenPRL2020, Jin_2022, Gao2019}, and in-plane magnetized systems hosting bimerons generate smaller stray fields.

In this article, we propose a model to stabilize a composite texture consisting of an antivortex and crescent-shaped vortex, see Fig.~\ref{fig1}(a), which is an in-plane analog of antiskyrmion~\cite{Koshibae2016, Nayak2017, Potkina2020}, and therefore dubbed as an asymmetric antibimeron.   We demonstrate that the AAB can be stabilized in a chiral ferromagnet with $D_{2d}$ symmetry and investigate its internal structure and current-induced dynamics. Moreover, we show that AABs with opposite topological charges can coexist in the same film. 
Thus, a magnetic system with AABs offers three topologically distinct states, allowing for their clear differentiation, and therefore realizing
a prototypical platform for ternary logic \footnote{See Supplemental Material at [URL will be inserted by publisher] for more details on ternary logic.}. Multi-valued logic systems have recently received significant attention as an approach to develop a more energy-efficient and higher-density electronics~\cite{Andreev2022, Yoo2021}. The ternary logic specifically is the most cost-effective when it comes to the use of hardware resources \cite{Hurst1984, Sandhie2022}. This logic can be achieved by employing magnetic system with spin textures of three different topological charges. While symmetric bimerons, skyrmions, and antiskyrmions cannot provide this functionality, AABs can. Since these three states are topologically protected, it also reduces the potential for errors. 
To efficiently operate this logic, we demonstrate that the total topological charge of the system can be tuned in AAB collisions by applying current in different directions. Furthermore, we extend Thiele’s equation approach to a system of two elliptic merons comprising the AAB, enabling us to accurately describe the current-driven dynamics of AABs in excellent agreement with our micromagnetic simulation results.

{\it Model.--}
We consider a ferromagnetic film with the exchange interaction, uniaxial anisotropy, DMI, and Zeeman coupling.  Assuming slowly varying magnetization, the magnetic free energy of the system can be written as 
\begin{equation}
F[\mathbf{m}]=\int d^3r \left[A(\mathbf{\nabla} \mathbf{m})^2 + \epsilon_{a} + \	\epsilon_{\rm{DM}} - M_{s}\mathbf{m}\cdot\mathbf{B} \right],
\label{eq1}
\end{equation}
where $A > 0$ is the exchange constant, $\mathbf{m}$ is the normalized magnetization,  $\epsilon_{a}=K_{x}[1-(\mathbf{m}\cdot  \mathbf{e}_{x})^2]$ is the in-plane easy-axis anisotropy energy density with anisotropy constant $K_{x}>0$, $\epsilon_{\rm{DM}}$ is the DMI energy density, $\mathbf{B} = B_{x} \mathbf{e}_{x}$ is the external magnetic field, and $M_{s}$ is the saturation magnetization. We consider an experimentally relevant DMI found in antiskyrmion systems, such as Heusler compounds $\mathrm{Mn_{1.4}Pt_{0.9}Pd_{0.1}Sn}$ \cite{Nayak2017}, $\mathrm{Mn_{2}Rh_{0.95}Ir_{0.05}Sn}$ \cite{Jena2020}, and $\mathrm{Mn_{1.4}PtSn}$~\cite{He2024}, which takes the form:
\begin{equation}
	\epsilon_{\rm{DM}} = D \left(\mathbf{e}_x \cdot \mathbf{m} \times \partial_y\mathbf{m} + \mathbf{e}_y \cdot \mathbf{m} \times \partial_x\mathbf{m}\right),
	\label{eq2}
\end{equation}
where $D$ is the DMI constant, and the DMI vectors are depicted in Fig.~\ref{fig1}(b) by the red arrows; this type of DMI is allowed under the $D_{2d}$ crystallographic point group of the host lattice. Beyond Heusler compounds, $D_{2d}$-symmetric $MX_2$ monolayers, including $\mathrm{NbO_2}$, $\mathrm{NbS_2}$, $\mathrm{NbSe_2}$, $\mathrm{MoS_2}$, and $\mathrm{MoSe_2}$, have also been theoretically proposed to host anisotropic DMI enabling antiskyrmion stabilization~\cite{Zhang2026}. 

We emphasize that our model utilizes an in-plane easy-axis anisotropy because it yields the largest parameter space over which AABs remain stable, thereby greatly facilitating a systematic study. However, this specific form of anisotropy is not strictly required. We find that AABs can also be stabilized in the complete absence of magnetic anisotropy, as well as in materials featuring a perpendicular out-of-plane anisotropy, $\epsilon_{a}=K_{z}[1-(\mathbf{m}\cdot \mathbf{e}_{z})^2]$. In the latter scenarios, however, the system must remain magnetized in plane, which necessitates an appropriately tuned external field $B_x$~\footnote{See Sec. II of Supplemental Material at [url will be inserted by publisher] for details on the micromagnetic simulations.}. More generally, in analogy to other topological spin textures, AABs can in principle emerge within different magnetic backgrounds, such as stripe or conical phases, which would inherently involve different parameter ranges and yield altered texture properties. In this work, we focus on AABs as topologically nontrivial states stabilized on a uniform ferromagnetic background. In materials that typically host antiskyrmions, this in-plane background can be readily achieved either by leveraging the dominant in-plane shape anisotropy in thin films or by applying a sufficient in-plane magnetic field.

By numerical minimization~\cite{OOMMF, Beg2022} of the free energy functional, Eq.~\eqref{eq1}, we obtain an equilibrium magnetic texture shown in Fig.~\ref{fig1}(a). This texture is not the ground state~\footnote{We note that the subsequent analysis is performed for the zero-temperature case to examine the intrinsic properties of the system. Crucially, the stability of AABs also persists at finite temperatures, and their robustness against thermal fluctuations is explicitly demonstrated in Sec. VI of the Supplemental Material at [url will be inserted by publisher], which includes Refs.~\cite{Garanin1997, Evans2012,Vervelaki2024,Pizzochero2020,Gong2017,Huang2017_2}.}, but a lowest excited nontrivial topological state of a system described by free energy $F$ in a certain window of material parameters~\cite{Suppl4, Suppl3}.
AABs are asymmetric spin textures consisting of an antivortex and a crescent-shaped vortex (red and blue regions in Fig.~\ref{fig1}(a), respectively). By comparing with a symmetric bimeron~\cite{Gobel2019}, shown in Fig.~\ref{fig1}(d) and having a different DMI [Fig.~\ref{fig1}(e)], where vortex and antivortex have the same arched shape, the antivortex in the AAB is more elliptically symmetric and plays a dominant role  as being more energetically favorable for the DMI given by Eq.~\eqref{eq2}, and thus this texture is called an antibimeron.

To explain the asymmetric shape of the AAB we analyze the influence of the DMI [Eq.~\eqref{eq2}] on the spin texture, as among the energy terms in Eq.~\eqref{eq1} only the DMI energetically favors out-of-plane spin canting, transforming the uniform magnetic state into a swirling one~\cite{Suppl2}.
This DMI prefers the formation of an antivortex, which has lower energy than a vortex. However, as a single antivortex cannot be stabilized by itself in a uniform in-plane FM background due to the topological constraints, the formation of a vortex with the opposite polarization becomes necessary. This leads to the creation of a coupled vortex-antivortex pair. Since a symmetric vortex has higher energy, it is less stable than the antivortex and undergoes deformation to minimize the overall energy of the AAB. Specifically, the vortex wraps around the antivortex in a crescent-like configuration, giving rise to the asymmetric shape of the AAB~\cite{Suppl3}.

\begin{figure}[tbp]
	\centering
	\includegraphics[width=1\linewidth]{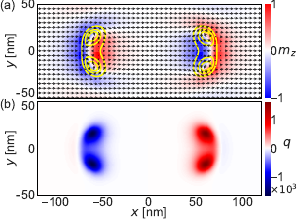}
	\caption{(a) Magnetization profiles of AABs with opposite topological charges stabilized within the same film. 
	The yellow lines depict contours of topological charge density. The color indicates the magnitude of  $m_{z}$. (b) The corresponding topological charge density distribution $q$.}
	\label{fig2}
\end{figure}

AABs exhibit intriguing topological properties. In general, a nontrivial spin texture is characterized by its topological charge:
\begin{equation}
Q = \int\! q dxdy, \qquad q=\frac{1}{4\pi} \mathbf{m}\cdot\left(\frac{\partial \mathbf{m}}{\partial x} \times \frac{\partial \mathbf{m}}{\partial y}\right),
\label{eq5}
\end{equation}
where $q$ is the topological charge density. As shown in Fig.~\ref{fig1}(c), $q$ of the AAB displays an asymmetric distribution with two distinct peaks. This contrasts sharply with the symmetric ring-shaped distributions observed in skyrmions and symmetric bimerons, see Fig.~\ref{fig1}(f). A unique feature of AABs is that configurations with $Q=+1$ and 
$Q=-1$ possess the same energy, allowing them to coexist simultaneously within the same film, as we demonstrate in Fig.~\ref{fig2}. These AABs with opposite topological charges are related by a mirror transformation with respect to the $yz$-plane, $\mathcal{M}_x\!\!: \mathcal{M}_x \mathbf{m} = [m_x(-x, y), -m_y(-x, y), -m_z(-x, y)]$. Given that the gradient transforms as $\mathcal{M}_x \mathbf{\nabla} = [-\partial_x, \partial_y ]$, the DMI in Eq.~\eqref{eq2} is symmetric under this mirror transformation. One can check that all other terms in the free energy, Eq.~\eqref{eq1}, are also symmetric under this transformation, and thus AABs with opposite topological charges are energetically degenerate \footnote{Note that this is not the case for symmetric bimerons \cite{Gobel2019}, whose DMI is given by $D (\mathbf{e}_y \cdot \mathbf{m} \times \partial_x\mathbf{m} +\mathbf{e}_z \cdot \mathbf{m} \times \partial_y\mathbf{m})$ and the vectors are depicted in Fig.~\ref{fig1}(e) by red arrows. Since this DMI is antisymmetric under the same mirror transformation, the symmetric bimerons favor a single topological charge value.}. 
This is in stark contrast to symmetric textures, such as skyrmions, antiskyrmions, and symmetric bimerons, where the background magnetization is perpendicular to all DMI vectors; for instance, when a skyrmion is stabilized, an antiskyrmion is energetically unstable.
The degeneracy between AABs with $Q=\pm 1$ is lifted when a magnetic field perpendicular to all DMI vectors (i.e., along $z$-axis) is applied.
As the field strength increases, the energy difference between AABs of opposite topological charges grows, eventually destabilizing one of the configurations. Our micromagnetic simulations reveal a critical field of $|B_z| \approx 45$ mT for this destabilization. The sign of $B_z$ unambiguously controls the sign of the stabilized topological charge $Q$. This property enables selective control of the total topological charge in the system using perpendicular magnetic fields. The coexistence of spin textures with $Q=+1$ and $Q=-1$ within the same material offers exciting possibilities for ternary devices, where three logic states--“0” (no AAB), “+1,” and “-1”--can be utilized. This functionality could be harnessed for topological computing applications~\cite{Li20, Zhang2015, Xia2022}.

\begin{figure}[tbp]
	\centering
	\includegraphics[width=1\linewidth]{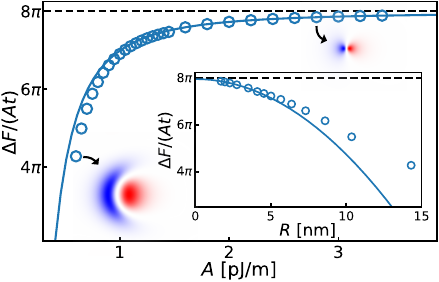}
	\caption{The formation energy of the AAB, $\Delta F$, as a function of the exchange constant $A$. The inset shows the relationship between the AAB's energy and its characteristic size $R$. Blue circles represent numerical data, while blue curves show fits to \(\Delta F/(At)=a+b/A^2\) in the main panel and to \(\Delta F/(At)=a+bR^2\) in the inset, where \(a\) and \(b\) are fitting parameters. 
	}
	\label{fig3}
\end{figure}

Next, we discuss the formation energy of AABs. Similar to skyrmions, an isolated AAB is an excited state in chiral in-plane magnets. As shown in Fig.~\ref{fig3}, the formation energy of the AAB relative to the ground state (uniform magnetization) asymptotically approaches the Belavin-Polyakov exchange limit~\cite{Belavin1975, Tretiakov2007, Bachmann2023}, $\Delta F \to 8\pi At$, with $t$ being the film thickness, as the exchange constant $A$ increases. This behavior arises because an increase in $A$ (with all other parameters held constant) leads to smaller AABs with larger rotational angles between neighboring magnetic moments. Since the exchange energy scales with the square of the magnetization gradient, while other energy terms exhibit weaker dependencies on spatial magnetization gradients, the exchange contribution becomes dominant in the limit of small AABs. The next significant contribution to the AAB formation energy in the small AAB limit comes from the DMI, as it is linear in the magnetization gradient. We numerically verified that, as the AAB size increases, the DMI gives a negative contribution to the AAB formation energy  $\Delta F$. The inset shows $\Delta F $ dependence on AAB's characteristic size $R$. Due to the asymmetric shape of the AAB, we define this size as $R=\sqrt{S}$, where $S$ is the area enclosed by the $m_x = 0$ contour~\footnote{See Fig.~S2 in Sec.~II of Supplemental Material at [url will be inserted by publisher]}. Using the relationship between $R$ and $A$~\cite{Suppl3}, the inset of Fig.~\ref{fig3} reveals that the AAB formation energy depends approximately linearly on $R$ for small AABs.  The detailed effect of other energy terms on the AAB size is described in the Supplemental Material~\cite{Suppl3}.

\textit{Current-driven motion of AABs.--}  To study the current-driven dynamics of AABs, we employ the Landau-Lifshitz-Gilbert (LLG) equation describing magnetization evolution under the influence of the spin-transfer torques \cite{Li2004, Thiaville2005}, 
\begin{equation}
\frac{d\vm}{dt} = \gamma \mathbf{H}_{\rm eff} \times \vm + \alpha \vm \times \frac{d\vm}{dt} - \left(\mathbf{u} \cdot \mathbf{\nabla} \right) \vm + \beta \vm \times \left( \mathbf{u} \cdot \mathbf{\nabla} \right) \vm,
	\label{eq:LLG}
\end{equation}
where $\mathbf{H}_{\rm eff} = - (\mu_0 M_s)^{-1} \delta F[\mathbf{m}] / \delta \vm$ is the effective magnetic field, $\mu_0$ is the vacuum permeability,  $\gamma = g|e|/(2mc)$ is the gyromagnetic ratio, $\alpha$ is the Gilbert damping constant, $\beta$ is the nonadiabatic spin-transfer torque constant, and $\mathbf{u} = P g \mu_B \mathbf{j}_e/(2 e M_s) $ with $P$ being the electron's spin polarization, $g$ being the $g$-factor, $\mu_B$ being the Bohr magneton, and $\mathbf{j}_e$ being the current density.

\begin{figure}[tbp]
	\centering
	\includegraphics[width=1\linewidth]{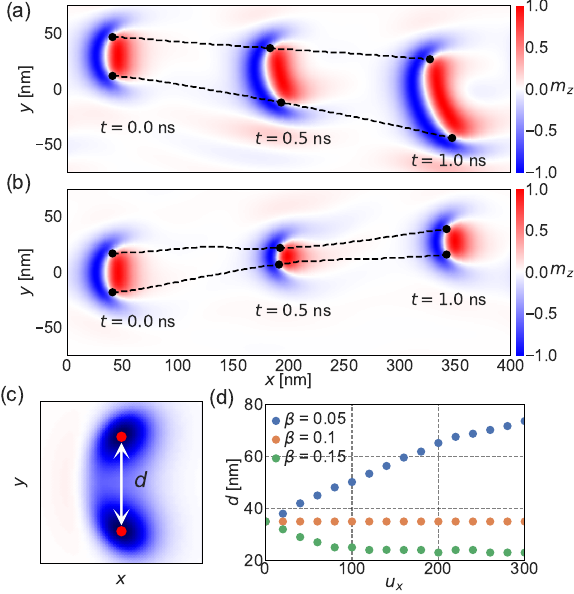}
	\caption{
			(a, b) The AAB trajectory under a DC current along the $x$-axis for $u_x = 200$ and $\alpha=0.1$ in case of (a) $\beta=0.05$ and (b) $\beta=0.15$. The dashed lines are the trajectories of the two peaks of topological charge density. Color indicates the $z$-component of magnetization. 
			(c) Topological charge density. The characteristic texture size $d$ is defined as the distance between the two peaks of the topological charge density (marked by red dots).
			(d) The size $d$ after a 1 ns current pulse as a function of current strength $u_x$ for~$\beta>\alpha$, $\beta=\alpha$, and $\beta<\alpha$ (here $\alpha=0.1$).
	}
	\label{fig4}
\end{figure}

We numerically solve the LLG equation with the current applied along $x$-direction ($\mathbf{u} = u_x \mathbf{e}_x$) to analyze the AAB motion.
As shown in Figs.~\ref{fig4}(a) and~\ref{fig4}(b), the spin-transfer torque induces not only the translational motion but also the deformation of the texture; the deformation trends depend on the $\alpha$ to $\beta$ ratio~\footnote{See Supplementary Movies 1 and 2 at [url will be inserted by publisher] for AAB dynamics with $u_x > 0$.}. 
When $\alpha > \beta$, the AAB elongates in the transverse direction and exhibits a counterclockwise tilt, in addition to moving with a negative transverse velocity, as illustrated in Fig.~\ref{fig4}(a).
The dashed lines in the figure represent the trajectories of the two peaks of the AAB’s topological charge density, marked as red dots in Fig.~\ref{fig4}(c).
Conversely, when $\alpha < \beta$, the AAB shrinks during its motion and exhibits a positive AAB Hall angle, as shown in Fig.~\ref{fig4}(b).
In this case, the distance between the two peaks, $d$, see Fig.~\ref{fig4}(c), oscillates over time, indicating a repulsive interaction when $d$ becomes small. 
Note that the AAB Hall angle is proportional to $(\alpha - \beta)$, which explains the sign change for the transverse motion.
For $\alpha = \beta$, the system becomes translationally invariant, and thus the AABs move without any deformation.

Figure~\ref{fig4}(d) shows the AAB size after a 1 ns pulse for various current strength $u_x$.
For $\alpha > \beta$, the AAB size increases linearly from its equilibrium value, indicating that the attractive interaction between the two merons, corresponding to the two peaks of $q$, is relatively weak when their separation $d$ exceeds the equilibrium distance.
For an infinite film, this leads to a continuous elongation of the AAB without annihilation in the long-time limit.
In contrast, for $\alpha < \beta$, the AAB size decreases linearly when the applied current is small ($u_x < 100 \ \rm m/s$) but saturates at higher currents due to the repulsive interaction between the merons 
\footnote{The potential structure is similar to interatomic potentials, such as Lennard-Jones potential.}. At sufficiently long times, the oscillations of $d$ are suppressed by Gilbert damping, and the distance between the two peaks of $q$ approaches a finite value.
Notably, reversing the current direction inverts the dependence on $\alpha-\beta$: When the current is applied in $-x$ direction, the AAB shrinks for $\alpha > \beta$, while it elongates for $\alpha < \beta$~\footnote{See Supplementary Movies 3 and 4 at [url will be inserted by publisher] for AAB dynamics with $u_x < 0$.}.

To understand the asymmetric dynamics of AABs, we apply the generalized Thiele’s equation approach~\cite{Thiele73, TretiakovPRL2008, Clarke2008}.
When a magnetic texture moves purely by translation without any deformations, $\vm(\vx, t) = \vm(\vx - \mathbf{v} t)$, its dynamics obeys the Thiele's equation,
$
	\hat{G}_{ij} \left(u_x \mathbf{e}_x - \mathbf{v} \right)_j + \hat{D}_{ij} \left( \beta u_x \mathbf{e}_x - \alpha \mathbf{v} \right)_j = 0,
$
where $\mathbf{v}$ is the velocity of the texture, $\hat{G}_{ij} = \int dx dy\ \vm \cdot \left( \partial_i \vm \times \partial_j \vm \right)$ is the gyrotropic tensor, and $\hat{D}_{ij} = \int dx dy\ \partial_i \vm \cdot \partial_j \vm$ is the dissipative tensor. Here, indices $i$ and $j$ run over the spatial coordinates $x$ and $y$.
Obviously, since AABs deform during motion, this standard Thiele’s approach does not fully describe their dynamics. However, if we focus separately on the dynamics of two individual merons, building blocks of the AAB, the motion can be approximated by their simple translation. 
These two merons are concentrated around the two peaks of topological density in the AAB, see e.g. Fig.~\ref{fig4}(c).
As seen from the topological charge density shown in Fig.~\ref{fig1}(c), each meron is elliptically deformed, rotated by approximately $\pm 45^\circ$, and carries a topological charge $|Q| = 1/2$.
In sharp contrast to symmetric spin textures (e.g., skyrmions or symmetric bimerons), whose $\hat{D}$ is diagonal, the off-diagonal components of the dissipative tensor, $D_{xy}$, are finite for elliptically deformed merons in the AAB.
Importantly, $D_{xy}$ is proportional to $\sin 2\delta$, where $\delta$ is the ellipse’s rotation angle, and crucially this means that the two merons in an AAB have opposite signs of $D_{xy}$~\footnote{See Supplementary material at [url will be inserted by publisher] for a detailed discussion on the off-diagonal component of the dissipative tensor.}.
Then, solving Thiele’s equation gives the meron velocities:
\bea
	v_x &=& \frac{16\pi^2 Q^2 + \alpha \beta \left( D^2 - D_{xy}^2 \right) - 4\pi (\alpha - \beta ) Q D_{xy}}{16\pi^2 Q^2 + \alpha^2 \left( D^2 - D_{xy}^2 \right)} u_x,
	\label{eq:v_parallel}
	\\
	v_y &=& \frac{4\pi ( \alpha - \beta) Q D}{16\pi^2 Q^2 + \alpha^2 \left( D^2 - D_{xy}^2 \right)} u_x,
	\label{eq:v_perp}
\eea
where $v_x$ and $v_y$ are the merons’ longitudinal and transverse velocities, respectively, and $D$ is the diagonal component of the dissipative tensor.
Equation~\eqref{eq:v_parallel} shows that due to finite $D_{xy}$, the two merons in an AAB move with different longitudinal velocities, causing elongation and rotation of the texture. Furthermore, the transverse velocity [Eq.~\eqref{eq:v_perp}] is proportional to $(\alpha - \beta)u_x$, leading to a sign change depending on the $\alpha/\beta$ ratio and sign of $Q$, which is in excellent agreement with our numerical results
\footnote{Note that this consideration neglects the interaction between the merons and as a result fails to capture some details, such as saturation in the size when the texture shrinks and the difference in the Hall velocities, see Fig.~\ref{fig4} (a) and (b).
Therefore, a more detailed theory, which includes the effect of interactions, would be required to describe this dynamics fully.}.

{\it AAB collisions.--}
Since AABs with opposite topological charges can coexist in the same film, their collisions present an interesting phenomenon. To investigate these collisions numerically, we drive AABs by current utilizing spin-transfer torques [Eq.~\eqref{eq:LLG}] and leverage the fact that the transverse component of AAB motion depends on the sign of $Q$ due to the antibimeron Hall effect as shown by Eq.~\eqref{eq:v_perp}. By appropriately positioning the initial locations of AABs, we can move them toward each other. In practice, creation of AABs can be realized by leveraging techniques similar to those used for skyrmion nucleation~\cite{Buttner2017, Kern2022, Kurebayashi2022}.

First, we examine collisions between the AABs of opposite charge by applying a current along $y$-axis~\footnote{See Supplementary Movies 5 and 6 at [url will be inserted by publisher] for AAB collisions with $u_y > 0$ and $u_y < 0$, respectively.}. From the perspective of topological charge conservation, pair annihilation is expected upon collision. In this case, the size of AABs with opposite $Q$ remains the same during the current-driven motion. As a result, AABs annihilate with each other, thus conserving the total topological charge as expected, while the energy released in the collision is carried away by the spin waves. Moreover, we identify a critical behavior in this type of collisions. When the applied current is small, the two AABs do not annihilate but instead move alongside each other along the current~\footnote{See Supplementary Movie 7 at [url will be inserted by publisher] for behavior in the small current regime.}, due to their repulsive interaction stemming from their rigidity. In contrast, for the collisions when the current is applied along the $x$-axis~\footnote{See Supplementary Movies 8 and 9 at [url will be inserted by publisher] for AAB collisions with $u_x > 0$ and $u_x < 0$, respectively.}, the AAB sizes evolve asymmetrically during motion: One AAB expands, while the other shrinks, depending on the current direction and the $\alpha/\beta$ ratio. Since smaller AABs are less stable (c.f.~Fig.~\ref{fig3}), unlike the $y$-axis case, the larger AAB always survives after the collision.
Consequently, topological charge is not necessarily conserved in AAB collisions, enabling controlled manipulation of the total $Q$ in a film with multiple AABs. Thus, the overall topological charge of the AAB system can be electrically tuned by applying currents in different directions.

{\it Conclusions.--}
We theoretically propose a model to stabilize asymmetric antibimerons -- composite spin textures consisting of an antivortex and a crescent-shaped vortex -- and investigate their equilibrium and dynamical properties.
We show that AABs can be stabilized in a ferromagnet with an in-plane uniaxial anisotropy and $D_{2d}$-symmetric DMI. Since some Heusler compounds are known to host such DMI~\cite{Nayak2017, Jena2020}, we anticipate that AABs will be experimentally observed in the near future. Moreover, we demonstrate the coexistence of AABs with opposite topological charges in the same film, in contrast to skyrmionic systems, where the sign of the topological charge is uniquely determined. Furthermore, we show -- both analytically and through micromagnetic simulations -- that the current-induced dynamics of AABs exhibit anisotropic behavior depending on the applied current direction. This behavior is explained by extending Thiele’s equation approach to a system of two elliptic merons comprising the AAB. We also demonstrate that the total topological charge in AAB systems can be controlled by static magnetic fields perpendicular to the film plane or independently via AAB collisions, achieved by selectively applying electric current along or perpendicular to the magnetic anisotropy direction. Ultimately, AABs offer a promising avenue for the realization of spintronic ternary memory and logic. Finally, we note that our results for AABs are fully applicable to other asymmetric spin textures stabilized by different types of DMIs, such as asymmetric bimerons~\cite{Shen2020, Ohara2022, Mukai2024, Babu2024, Yu2024} and asymmetric (anti)skyrmions.

\textit{Acknowledgments.--} O.A.T. acknowledges the support from the Australian Research Council (Grants No.~DP200101027 and No.~DP240101062) and the NCMAS grant.

\textit{Data availability.--} There are no publicly available research data or software supporting this manuscript. Requests for further information or data should be sent to the authors.

P.A.V. and D.K. contributed equally to this work. O.A.T. conceived and supervised the project. P.A.V. and D.K. conducted simulations on the static properties, while D.K. conducted simulations on the dynamical properties and constructed the analytical theory with the guidance of O.A.T. All authors contributed to discussions and writing the manuscript.

\bibliography{refs}

\end{document}